\documentclass[letterpaper,twocolumn,10pt]{article}
\usepackage{usenix-2020-09}

\usepackage{amsmath}
\usepackage{graphicx}
\usepackage{xspace}
\usepackage{xcolor}
\usepackage{soul}
\usepackage{adjustbox}
\usepackage{tablefootnote}
\usepackage{url}
\usepackage{tabularx,colortbl}
\usepackage{multirow}
\usepackage{lscape}
\usepackage{algorithm}
\usepackage{algpseudocode}
\usepackage{mathtools}
\usepackage{url}
\usepackage{lipsum}
\usepackage{amssymb}
\usepackage{booktabs}
\usepackage{enumitem}
\graphicspath{{Images/}}
\usepackage{filecontents}
\usepackage{comment}

\newcommand{\method}{IC-SECURE\xspace}

\begin{document}
%-------------------------------------------------------------------------------

%don't want date printed
\date{}

\title{\Large \bf \textit{\method}: Intelligent System for Assisting Security Experts in Generating Playbooks for Automated Incident Response}

%for single author (just remove % characters)
\author{
{\rm Ryuta Kremer, Prasanna N. Wudali, Yuval Elovici, Asaf Shabtai}\\
Department of Software and Information Systems Engineering \\
Ben-Gurion University of the Negev
\and
{\rm Satoru Momiyama, Toshinori Araki, Jun Furukawa}\\
 NEC Corporation
% copy the following lines to add more authors
% \and
% {\rm Name}\\
%Name Institution
} % end author

\maketitle

%------------------------------------------------
\begin{abstract}

Security orchestration, automation, and response (SOAR) systems ingest alerts from security information and event management (SIEM) system, and then trigger relevant playbooks that automate and orchestrate the execution of a sequence of security activities. % (i.e., workflows). 
SOAR systems have two major limitations: \textit{(i)} security analysts need to define, create and change playbooks manually, and \textit{(ii)} the choice between multiple playbooks that could be triggered is based on rules 
defined by security analysts.
To address these limitations, recent studies in the field of artificial intelligence for cybersecurity suggested the task of interactive playbook creation.
In this paper, we propose \textit{\method}, an interactive playbook creation solution based on a novel deep learning-based approach that provides recommendations to security analysts during the playbook creation process.
\method captures the context in the form of alert data and current status of incomplete playbook, required to make reasonable recommendation for next module that should be included in the new playbook being created.
We created three evaluation datasets, each of which involved a combination of a set of alert rules and a set of playbooks from a SOAR platform. 
We evaluated \method under various settings, and compared our results with two state-of-the-art recommender system methods. In our evaluation \method demonstrated superior performance compared to other methods by consistently recommending the correct security module, achieving $precision@1>0.8$ and $recall@3>0.92$.

\end{abstract}

\section{\label{sec:intro}Introduction}

To reduce security analysts' workload and improve their efficiency, many organizations use security orchestration, automation, and response (SOAR) systems in their security operation centers (SOCs)~\cite{bridges2023testing}.
SOAR solutions are designed to integrate a wide range of security tools, streamline various SOC processes, and automate disparate security incident response tasks. 

SOAR systems operate in conjunction with security incident and event management (SIEM) systems~\cite{kinyua2021ai}, which are responsible for detecting malicious or suspicious activities in the raw logs of various sensors and raising alerts when specified signatures or anomalies are detected. 
These alerts are then forwarded to the SOAR system, which triggers investigative (e.g., check for URL reputation or get user/asset information) and response (e.g., block IP in firewall or kill a malicious process on an endpoint) activities.

In SOAR systems, a sequence of activities is carried out based on \emph{playbooks} manually defined and created by the security analysts.
These playbooks are developed in accordance with incident response plan (IRP) policy documents, which outlines the step-by-step activities for guiding security analysts in investigating and responding to alerts.

\begin{figure*}[htbp]
    \centering
    \includegraphics[width=0.90\linewidth,trim={0cm 1.4cm 0cm 0.2cm},clip]{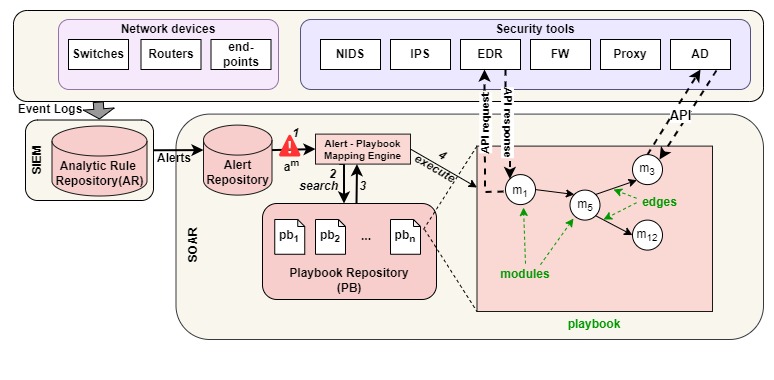}
    \caption{Overview of a SOAR system as it executes a playbook.}
    \label{fig:SOAR}
\end{figure*}

There are two main types of playbooks: ($i$) investigative playbooks, and ($ii$) response playbooks.
\textit{Investigative playbooks} orchestrate the steps involved in collecting additional contextual information about an alert.
Response actions are codified in the form of \textit{response playbooks}, which orchestrate the steps required to mitigate a known or expected security event, incident, or threat and prevent it from harming the entire network.
SOAR systems integrate and orchestrate the use of various security tools to perform these tasks.
As seen in Figure~\ref{fig:SOAR}, a playbook can be represented as a graph in which the nodes are modules and the edges are the direction of the logic flow and related conditions.

% [trim={left bottom right top},clip]

A \emph{module} in a playbook is the specific configuration of a security tool in the SOAR system.
Modules link security tools to the SOAR platform, with the help of application programming interfaces (APIs), and are responsible for performing actions (investigative or response) to address a security alert.
For example, the \textit{Checkpoint} security module links the Checkpoint firewall deployed in the organization's network to the SOAR system. 
This security module can perform various actions related to the Checkpoint firewall, such as adding a new rule to the firewall to block traffic to an IP address outside the organization's network (e.g., when an alert for a suspicious IP address is raised).

There is a wide range of security tools available today.
Defining and implementing playbooks that are aligned with an organization's IRPs as well as integrating the relevant state-of-the-art security tools is a challenging task requiring specific skills and expertise, both of which organizations often lack~\cite{sworna2023apiro,sworna2023irp2api}.
This is compounded by the fact that security analysts are typically faced with thousands of alerts on a daily basis~\cite{kinyua2021ai} and hence, they do not have time to tend to playbooks. 

There are three main challenges associated with SOAR system. First, when new security tools (i.e., modules) are added, security analysts must create new playbooks or update, and maintain the existing playbooks for known alerts~\cite{sworna2023irp2api}.
To do this, security analysts must have significant expertise in handling security incidents~\cite{bridges2023testing}, as well as in-depth knowledge on the various security tools integrated in SOAR systems and the actions each tool can perform~\cite{sworna2023apiro,sworna2023irp2api}.
Second, once defined, playbooks are hard-coded for a fixed set of alerts and are fairly static and rigid~\cite{sumologic}.
This may be acceptable in the case of investigative playbooks, which may not need to be changed frequently~\cite{bridges2023testing}, but it is less desirable in the case of response playbooks, which may need to be changed in order to adapt to emerging threats and novel, previously unseen alerts.
Consequently, the response playbooks in the SOAR system repository may rendered useless against novel alerts.
Therefore, when faced with such alerts, there is a need to quickly create a new playbook~\cite{ibm_soar}.
Third, in most SOAR systems, rule-based methods are employed to map playbooks to alerts~\cite{kraeva2021application}.
The main disadvantage of rule-based SOAR systems is that a playbook's relevance to an alert depends on the rules rather than the context of the alert.
Moreover, it is difficult to validate and update the rules and ensure their completeness and correctness.
Given these challenges and the time it takes to both create new playbooks and resolve the errors in newly created playbooks, security analysts require assistance in easily and quickly creating, updating, and maintaining playbooks and rules associated with them.

Previous works in the domain of response recommendation for security alerts proposed heuristic-based methods for automated response selection, which are based on penalty cost functions representing the effectiveness of the selected response~\cite{foo2005adepts,1176302}.
These methods, however, are inflexible as they cannot easily adapt to the changes in the environment, less accurate as the response selected depends on the calculated cost, lack context, and lack transparency as such models behave mostly as black boxes due to their complexity.
Kraeva et al.~\cite{kraeva2021application} proposed a deep learning-based method for recommending the most relevant playbook for handling an incident.
However, since the purpose of their proposed model is the automated selection of a playbook, it is more suitable for alert triaging tasks where the investigative playbooks are fairly static and workload reduction task for security analysts where response playbooks triggered for \emph{known} alerts have predefined IRPs and is less suitable for novel alerts for which existing IRPs are not suitable.
Two similar frameworks, APIRO~\cite{sworna2023apiro} and IRP2API~\cite{sworna2023irp2api}, were proposed to automate security module recommendations for the actions mentioned in the IRPs.
Such methods, however, do not capture the context provided by  the alert that should be handled and prior security modules, which is crucial in order to recommend the module appropriate to carry out the next action defined in the IRP.
Islam et al.~\cite{islam2019ontology} proposed an ontology-driven approach \textit{OnSOAP} for the automated selection and execution of an IRP based on the alert type.
The main drawback of this approach is that it requires a detailed definition of alert to IRP mappings. 
These ontologies store information regarding the various relationships required by SOAR (for example, functional and non-functional activities performed by security systems), and IRP definitions. 
Thus, it is not suitable for handling novel alerts.
While these studies have addressed the aspects of automated execution of IRPs and selection of appropriate security modules for actions mentioned in the IRPs, the proposed methods are limited for creating playbooks, and moreover, the studies did not present a full-scale recommender system capable of guiding a user through incident handling and suggesting which steps to take~\cite{husak2022sok}. 

In this paper, we introduce \emph{\method}, an intelligent context-based security module recommender system for playbook creation; based on a novel deep learning-based model, designed to provide continuous support to security analysts responsible for generating new playbooks for previously unseen security alerts.
In \method, a deep learning (DL) model is used as a recommender system; by providing the model with the context in the form of information (features) about the alert and current state of the playbook being implemented in graph representation (i.e., playbook subgraph) along with the information about the propagating node with the aim of providing relevant module recommendations for the next module in the playbook.
In this research, playbooks are represented as directed graphs with modules in the playbook represented by the nodes, and the direction of logic flow and conditions associated with the flow are represented as edges.

We evaluate \method using three datasets, each of which includes alert rules with predefined playbooks triggered by these alerts, and examine our DL model's ability to provide relevant recommendations for security modules. 
The results show that our method is capable of providing reasonable recommendations for a new node in the playbook, i.e., reconstruct links between nodes in the original playbook.
Specifically, in our evaluation \method demonstrated superior performance compared to other methods by consistently recommending the correct security module, achieving $precision@1>0.8$ and $recall@3>0.92$.

The main contributions of our paper are: 
\begin{itemize}[noitemsep, nolistsep]
    \item We propose unique representations of alerts and playbooks as constructive and meaningful input to the deep learning-based model. 
    These representations provide context to the model, enabling it to produce reasonable recommendations.
    \item We propose a novel recommender system model to recommend security modules that should be included in the playbook during the playbook's design.
    \item We collected three unique labelled datasets of alerts and mapped playbooks, which we have made available to the SOAR system research community.
\end{itemize}

\section{\label{sec:related}Related Work}

Previous works in the domain of response recommendation for security alerts proposed ontology, heuristic, and machine learning (ML)-based methods for automated response selection.
A summary of these works is presented in Table~\ref{tab:literature_review}.

Souissi et al.~\cite{souissi} suggested an ML-based decision-making system that selects a defense mechanism for known alert types (i.e., attack classes). 
The authors categorized the defense mechanisms as follows: detection, prevention, mitigation, remediation, tolerance, and awareness. 
A supervised ML classifier model is trained on a dataset that maps attack classes to defense mechanisms.
This method, however, fails at suggesting a defense mechanism for novel alert types.
Also, this method can only answers the question "\textit{what needs to be done?}" and fails to answer the question "\textit{How it needs to be done?}"

Some intrusion response systems proposed in other studies have attempted to incorporate automated response selection to alerts by calculating various metric values and then use simple heuristics to decide which response to use.
Toth et al.~\cite{1176302} proposed IE-IRS, an automated response selection model that calculates the penalty cost of each response action and selects the response action with the lowest penalty cost.  
Foo et al.~\cite{foo2005adepts} introduced ADEPTS, an adaptive intrusion response model that automates response selection based on the response index (RI) value.
This RI is determined by evaluating the response's effectiveness against a specific attack detected in the past and the potential impact, in terms of disruption, on legitimate users.
The response with the highest RI value is selected and deployed. 
In other studies that proposed automating the response selection process for intrusion response systems~\cite{balepin2003using,tanachaiwiwat2002adaptive,papadaki2006achieving,Stakhanova2007Cost,Kheir2010cost,4299772,5254241}, the authors defined various metrics in order to determine the appropriate response mechanism. 
Such heuristic-based cost calculation methods lack transparency and are inflexible, inaccurate and slow.

An ontology-based model for security orchestration platform was proposed by Chadni et al.~\cite{8812856}. 
The authors automate the incident response process by triggering the response activities according to the selected incident response plan (IRP). 
In this case, alerts are categorized based on the alert type, and the alert type to IRP mapping rules are defined based on security and business considerations and stored in the IRP repository. 
Each IRP contains a set of activities performed by security systems in response to a security alert.
These activities are classified as functional activities or non-functional activities.
The relationship between a security system and the activities it can perform is maintained in an ontology. 
The main drawback of this approach is the need for alert type to IRP mapping rules, ontologies illustrating various relationships related to the system, and IRP definitions.
If these requirements are not met, the model cannot perform properly.

Irina et al.~\cite{kraeva2021application} proposed a metric learning-based approach for automated playbook selection.
Their trained neural network maps the entire set of alerts into a vector space, in which alerts that are handled by same playbooks are closer to each other than to alerts handled by different playbooks.
For an given alert the model either finds a set of relevant playbooks from the repository of playbooks or informs the user that no suitable playbook exists in the playbook repository.
This model's main disadvantage is that it is only suitable for alert triaging tasks for which investigative playbooks recommended are predefined and are not frequently changed.
With response playbooks, which undergo frequent changes in order to adapt to novel alerts, this model fails to provide appropriate recommendations. 

APIRO~\cite{sworna2023apiro} is a framework for the automated recommendation of APIs in SOAR systems.
In APIRO, an API-specific word embedding model and convolutional neural network (CNN) model are used to predict the top-three relevant APIs for a natural language query submitted by a SOC analyst.
Similarly, IRP2API~\cite{sworna2023irp2api} automates the process of mapping cybersecurity IRPs to APIs of security tools.
It uses natural language processing (NLP) approaches to map each action in the IRP to a specific security tool API.
Both APIRO and IRP2API are reactive models that act as search engines to find relevant APIs to perform a particular task.
Such approaches do not capture the implicit or explicit context required to select appropriate response actions, and therefore they result in incorrect or no recommendations.

Martin et al.~\cite{husak2022sok} discussed the application of recommender systems in the incident handling and response process, and the challenges faced in their application in this setting. 
They considered the extent to which the phases and tiers of incident handling can be automated and assessed the maturity of existing approaches and tools.
The authors concluded their study by stating that they could not identify a a full-scale recommender system that would guide the user through incident handling and suggest which steps to take.
They also mentioned that the incident handling and response lifecycle is the longest and most strenuous phase and might face challenges such as scarce workforce and information overload to the security analysts.
The authors conclude the research by stating that these challenges can be addressed via decision support and recommender systems.

\begin{table*}[h!]
\caption{Summary of selected related work.}
\label{tab:literature_review}
\begin{adjustbox}{width=1.0\textwidth,center=\textwidth}
\begin{tabular}{@{}llllllll@{}}
\toprule
\textbf{Paper} & \textbf{Approach} & \textbf{Method} & \textbf{Dataset Used} & \textbf{Type of Data} & \textbf{Goal} \\ \midrule
Toth et al. (2002)~\cite{1176302} & Heuristics & Cost calculation & NA & NA &Cost-based response selection \\
Foo et al. (2005)~\cite{foo2005adepts} & Heuristics & Cost calculation & NA & NA & Cost-based response selection \\
Souissi et al. (2016)~\cite{souissi} & ML & Classification & custom & 432 pre-defined alerts & Incident classification \\ 
Chadni et al. (2019)~\cite{8812856} & Rules & Ontologies & custom & IRPs, rules, and ontologies & Automated execution of IRP\\
Irina et al. (2022)~\cite{kraeva2021application} & ML & Metric learning & VERIS\tablefootnote{https://verisframework.org/vcdb.html} and MITRE\tablefootnote{https://attack.mitre.org/} & Alerts and playbooks & Automated playbook selection \\
APIRO (2023)~\cite{sworna2023apiro} & ML & NLP and CNN & custom & API descriptions & Security tool API recommendation for IRPs \\
IRP2API (2023)~\cite{sworna2023irp2api} & ML & NLP & Cortex SOAR\tablefootnote{https://github.com/demisto/content}  & API descriptions and IRPs & Security tool API recommendation for IRPs\\

\bottomrule
\bottomrule

\method (our method) & ML & NN (NCF) & BOTS,\footnote[7]{https://github.com/splunk/botsv1 https://github.com/splunk/botsv2 \& https://github.com/splunk/botsv3} AD,\footnote[8]{https://github.com/splunk/attack\_data} and SSC\footnote[9]{https://github.com/splunk/security\_content}   & Alerts and playbooks & Security tool and its action recommendation \\
\bottomrule

\end{tabular}%
\end{adjustbox}
\end{table*}

\section{\label{sec:proposedmethod}Proposed Method}

In this section, we provide a detailed description of our proposed method - \method.
In our method, we follow an iterative process for crafting the new playbook. 
During each iteration, a security expert selects a node (i.e., security module) from the current partial playbook and augments it with an additional subsequent module. 
This process is repeated until no further modules are required for inclusion.

As illustrated in Figure~\ref{fig:NCFmodel}, our model takes as input (1) features of the alert for which we aim to generate the playbook, (2) the existing partial playbook, and (3) the current playbook node for which we intend to define the subsequent node. 
The output of our \method recommender system is a list of the top-\textit{k} recommended modules from which the security expert can select the next module.

The remainder of this section is organized as follows:
In Section~\ref{subsec:notations}, we provide the notations and definitions used in our proposed method.
In Section~\ref{subsec:high_level_arch}, we provide a high-level description of \method's architecture.

\subsection{\label{subsec:notations}Notations and Problem Statement}

\paragraph{Event logs.} Events are recorded by various devices (e.g., security tools, endpoints, and network devices), each of which may have a different structure.
These event logs are ingested by the SIEM system for storage, management, processing, and analysis. 
The SIEM system parses these logs into a common schema, which defines a set of keys $K = \{k_1, k_2,..., k_n\}$ to store information (i.e., key values) from the raw logs.
For example, Splunk SIEM uses the Common Information Model (CIM)~\cite{cdm,mehta2021data} to represent all the ingested logs, and it defines $\approx2,700$ keys at the time of publication of this research.

\noindent We denote the corpus of all the event logs ingested by the SIEM as $L$.
The information (attributes) of each event log $l \in L$ are represented in SIEM schema-compatible fields ($k_iv_i = <key_i>:<value_i>$) as \[l =\{k_1v_1,k_2v_2,...,k_pv_p\} \mid k_i \in K. \]  
Examples of $<key>:<value>$ pairs are: srcip: "172.16.22.10," dstip: "192.88.99.8," hostname: "DTP-17060," and process name: "nginx.exe."\\
Note that not all keys in $K$ are relevant for every event type. 
When a key $k_i \in K$ is irrelevant for a particular event, its value is set to $v_i=\varnothing$.

\paragraph{Alerts.} An alert is evoked in the SIEM system when a potential security threat is detected in the event logs. 
Generally, SIEM systems employ an analytical rule repository ($AR$) that searches through the parsed event logs to find suspicious activity, and all of the logs that match analytical rules ($ar$) are tagged as alerts.

\noindent We represent the analytical rule repository as a set of rules: $AR =\{ar_1,ar_2,...,ar_m\}$.
Each analytic rule can tag a single event or set of events as an alert.
We denote an alert $j$ that was triggered by the analytical rule $ar_m$ as: \[\{ar_m \mid l_1,l_2,...,l_t, \quad l_t  \in  L\} \Rightarrow a^{(m)}_j \] 
The information in each alert $a^{(m)}_j$ is represented using a subset of keys $K$, defined according to the logs $l$ that qualify as $a^{(m)}_j$. 
Therefore, we represent the alert data as: \[a^{(m)}_j = \{k_iv_i \mid k_i \in K_{ar_m}, K_{ar_m} \subset K \}\]

\paragraph{Modules.} 
SOAR systems integrate and orchestrate various security tools to enable them to work together.
A module ($m$) is a specific configuration of a security tool integrated in the SOAR system.
The corpus of tools integrated into a SOAR system is represented as, $M = \{m_1,m_2,\ldots, m_q\}$.
For example, $m$ can be Snort IDS, Checkpoint firewall, Crowdstrike EDR, or Symantec endpoint protection.
Each $m$ can perform single or multiple actions to provide protection against potential security attacks.
For example, $m$ can block ip, delete files, kill process, filter data, or format data.

\noindent A module is represented by a node in the playbook graph.
In addition, a module is linked to the corresponding security tool through the use of APIs, as presented in Figure~\ref{fig:SOAR}.
These APIs facilitate seamless communication and data exchange between the module and the underlying security tool, enabling the effective integration of diverse functionalities.

\noindent Throughout this paper, we refer to the module from which the recommendation is to be made as \textit{current node} (\textit{$m_t$}) and the new subsequent module that should be recommended as \textit{recommended node} (\textit{$m_{t+1}$}).

\paragraph{Playbooks.}
Each SOAR system maintains a repository of playbooks. 
We represent this repository ($PB$) as a set of playbooks ($pb$): $PB =\{pb_1,pb_2,...,pb_n\}$.
Each playbook is a graph in which the modules ($m$) serve as nodes, and the edges (i.e., links) represent the direction of the logical flow and the conditions associated with the flow.
Therefore, we represent each playbook ($pb$) as a directed graph with a set of nodes ($V$) and a set of edges ($E$):
\[pb =(V, E), \quad \forall pb \in PB, \]
where
\[V = \{m_1, m_2, \ldots, m_n\} \] 
\[E = \{e_1, e_2, \ldots, e_q\}, e_i = (m_{in}, m_{out}) \in V \times V.\]
When performing the module recommendation task, incomplete playbook graphs are created in every recommendation iteration.
We denote the current state of such incomplete playbooks as $pb^*$.

\paragraph{Alert playbook mapping.}
Existing rule-based SOAR systems trigger playbooks in response to alerts based on the mapping rules.
An alert can be mapped to a single playbook or a set of playbooks.
We represent these mapping rules as a consolidated function: \[f:AR\mapsto PB\] and use the mapping rules to create the dataset for our recommendation model.
Our model is trained to learn the relationships between alerts and playbooks, which enables it to generate relevant recommendations for modules (nodes).

\noindent For the task of playbook node recommendation, the top@$k$ recommendations for new nodes are the results of the recommender system. 
It is possible to have a situation in which no additional node should be added (i.e., the current explored node represents the last module to execute). 
Therefore, an artificial node (module) referred to as End-of-Playbook (EOP) can be recommended. 

In the course of standard SOC operations, situations may arise where the playbooks in the repository are unsuitable to response to a novel alert. 
Assuming that, following alert triaging and investigation, we have all the enriched information available for the novel alert, our method is designed to recommend components for a tailored playbook suitable for the novel alert.

\begin{figure*}[h]
    \centering
    \includegraphics[width=1\textwidth,trim={0cm 0.5cm 0cm 0cm},clip]{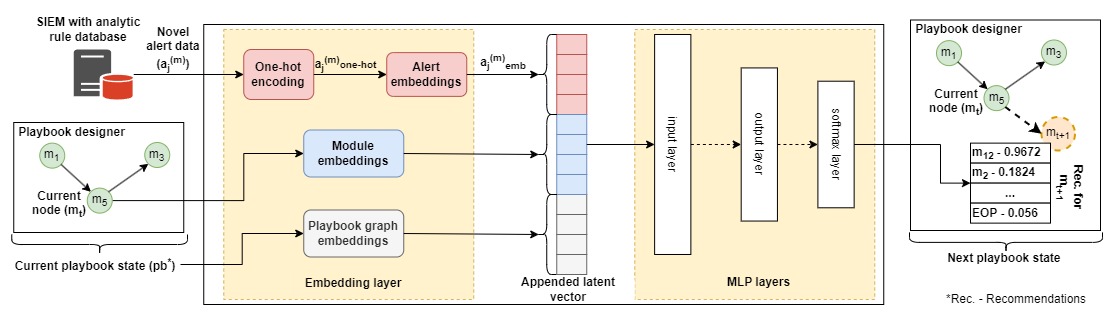}
    \vspace{-0.8cm}
    \caption{\method model overview.}
    \label{fig:NCFmodel}
\end{figure*}

\textit{\textbf{Problem Statement}:} Given context in the form of alert data from a novel alert $a^{(m)}_j$ triggered by an analytic rule (\textit{$ar_m$}), the current state of an incomplete playbook graph (\textit{$pb^*$}), and the information of the current node $m_t$, the recommender system should be able to provide the top-$k$ recommended modules from which the security analyst can select the next module $m_{t+1}$.
For example, in Figure~\ref{fig:NCFmodel} a new playbook is created for the novel alert $a^{(m)}_j$. 
The current state of the incomplete new playbook ($pb^*$) is $m_1 \rightarrow m_5 \rightarrow m_3$ and the current node is $m_5$. 
Based on this contextual information, the recommender system should be able to recommend $m_{12}$ in top@$k$ recommendations for the successor node of $m_5$ in the playbook graph.

\subsection{\label{subsec:high_level_arch}High-Level Architecture}

We propose \method, a contextual recommender system model based on neural collaborative filtering (NCF)~\cite{he2017neural} for recommending the next module $m_{t+1}$ to be included in the playbook being created (see Figure~\ref{fig:NCFmodel}).
The architecture consists of an one-hot alert encoding component, alert embedding component, module embeddings component for playbook nodes, playbook graph embeddings for playbook graph, and a deep learning (DL) model.

The alert encoder generates $a^{(m)}_{j,one-hot}$ for the $a^{(m)}_j$, and the alert embedding component of the system creates embeddings for alerts $a^{(m)}_{j,emb}$.
The module embedding component generates embeddings for current node ($m_t$) in the playbook graph. 
The playbook graph embedding component generates the embeddings for the current state of playbook graph $pb^*$ in the playbook designer.
Finally, the DL model combines these embeddings to calculate a score indicating the suitability of the next playbook node $m_{t+1}$ in the graph.
By learning the relationships between alerts and playbook nodes, the proposed model provides scores for recommendations during playbook generation, i.e., predicting new links from the current node to the new node during the playbook creation process.

\subsubsection{Alert Embedding}
In our proposed method, alert embedding plays a crucial role in capturing the essential information from the schema fields (i.e., $K$).
The utilization of alert embedding enables the efficient representation of alerts within a lower-dimensional embedding space, facilitating intelligent recommendations derived from these encoded representations.

The alert embedding process starts with the creation of a one-hot encoded vector for each alert.
An element in the vector has value of one if the corresponding key $k$ exists in the alert data.
Given an alert $a^{(m)}_j$, the one-hot encoded vector is represented as: 
\[
a^{(m)}_{j,one-hot} = \begin{cases}
    1 & k \in  K_{a^m} \\
    0 & \text{otherwise}
\end{cases}
\]

\begin{figure}[h]
    \centering
    \includegraphics[width=0.95\columnwidth,trim={0cm 1.0cm 0cm 0cm},clip]{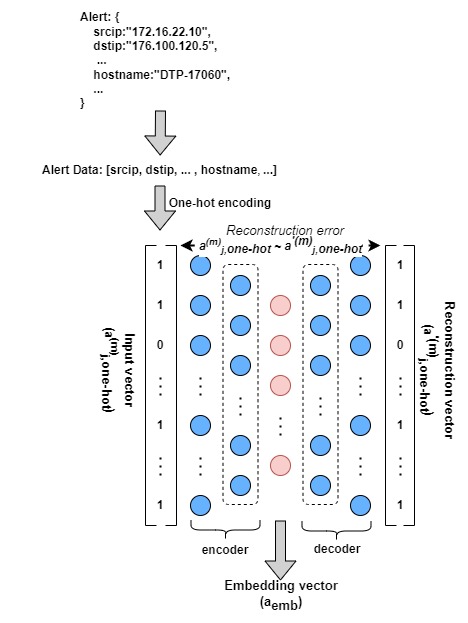}
    \caption{High-level overview of the alert embedding module.}
    \label{fig:alertembedding}
\end{figure}

In Figure~\ref{fig:alertembedding}, for example, the alert is represented in a SIEM schema that includes multiple keys ($k$), while three of the relevant keys for representing the alert are: \textit{srcip, dstip, and hostname}. 
Thus, the alert's one-hot vector representation would consist of a value of one for elements corresponding to \textit{srcip, dstip, hostname} along with other elements present in the alert and the remaining elements not present in the alert would be populated with a value of zero.

The one-hot vectors employed for alert representation are sparse, as only a limited set of keys is used to represent alerts.
Therefore, we use an autoencoder model for encoding the alert vectors into the lower-dimensional embedding vectors, $a^{(m)}_{j,emb}$.
The autoencoder is trained by minimizing the difference between the input vectors and reconstructed output vectors (i.e., $a^{(m)}_{j,one-hot} \approx a'^{(m)}_{j,one-hot}$).
The embedding vector $a^{(m)}_{j,emb}$ serves as a dense representation of the alert (capturing its important features) whose size is aligned with the embeddings of other elements.
Notably, the autoencoder is designed to handle unknown forms of alerts that may appear in the test data, ensuring its suitability for a wide range of alerts.

\vspace{-0.4cm}
\subsubsection{Module Embedding}
The objective of module embedding is to produce a numerical vector representation for each node ($m$), capturing the structural information of modules within playbook graphs.
This enables \method to recommend modules for the playbook being created which are relevant to the new node for which the link is propagating from the selected node.

\begin{figure}[h]
    \centering
    \includegraphics[width=0.98\columnwidth]{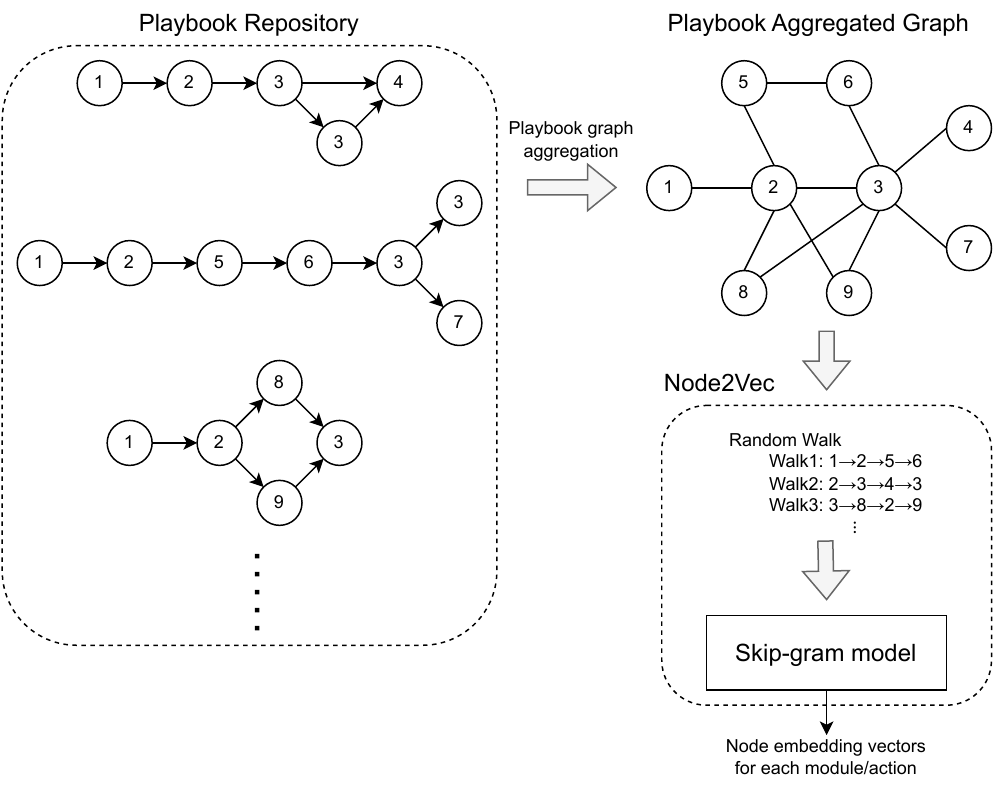}
    \vspace{-0.5cm}
    \caption{The security module embedding process, which is performed using node2vec~\cite{grover2016node2vec}.}
    \label{fig:node_embedding}
\end{figure}

The process of generating the module embedding is illustrated in Figure~\ref{fig:node_embedding}.
In the first step, the set of $PB = {pb_1, pb_2, ..., pb_n}$ available in our playbook repository is retrieved.
Then, using these playbooks, a unified playbook graph is generated, denoted as $pb_U$, where the nodes are all possible modules $V_U = V_{pb_1}\cup V_{pb_2}\cup ...V_{pb_n}$, and the edges are $E_U=E_{pb_1}\cup E_{pb_2}\cup ...E_{pb_n}$.
Intuitively, the set of nodes in the aggregated graph is the combined list of modules in all the playbooks in $PB$, and an edge between two modules $m_i$ and $m_j$ exists only if there is a link between them in at least one of the $pb \in PB$.
By aggregating the playbooks into a single graph, we aim to capture the relationships between modules.

Finally, the node2vec~\cite{grover2016node2vec} technique is used to derive a numerical vector representation for each $m_i$ in the unified graph.
This allows us to capture the collective influence of each module $m_i$ and dependencies with among other modules $m_j \in M$, facilitating the generation of meaningful embeddings.

\subsubsection{Playbook Graph Embedding}

One of the inputs for our model is the current state of the incomplete (partial) playbook, i.e., $pb^*$, for which we would like to recommend subsequent security modules. 
The graph embedding technique, specifically graph2vec~\cite{narayanan2017graph2vec}, enables us to represent $pb^*$ as a numerical vector, facilitating similarity comparisons between different graphs.

By capturing the structural characteristics and context of the $pb^*$ graph, graph embedding allows \method to generate meaningful representations that contribute to the generation of accurate and relevant recommendations of modules for playbook construction.

We consider two approaches for generating the playbook graph embedding: (1) capturing just the structure of the playbook graph, i.e., using the connections between modules while disregarding the type/name of modules (referred to as \emph{without attributes});
and (2) using a graph representation which includes the attributes of nodes, in our case, the type of security module (referred to as \emph{with attributes}). 
In the second case, multiple nodes with the same modules can be included in the graph representation with attributes.

\subsubsection{Security Module Recommender System}

Classical recommender system models have been widely used for various purposes, such as suggesting products in e-commerce~\cite{siva2014rec} and personalizing content feeds on social media~\cite{anandhan2018soc}.
These models typically rely on state-of-the-art techniques such as collaborative filtering~\cite{sarwar2001item,herlocker2004evaluating,zhang2016discrete,chen2017attentive} and content-based methods~\cite{lops2011content}, however, they often fail to capture complex and nonlinear relationships between users and items.
Recent advances in recommender system models that leverage deep learning techniques have led to enhanced performance.
These models can effectively model complex and nonlinear relationships between users and items, allowing for the representation of more sophisticated abstractions in the higher layers of the network~\cite{zhang2019deep}.
Furthermore, these methods have the capability to process diverse types of information related to users and items, encompassing contextual, textual, and visual data, by employing models like recurrent neural networks (RNNs) and convolutional neural networks (CNNs).

Therefore, for the recommender system, we used a deep learning (DL) model which performs the task of recommendation based on the embedding representation data.
We adapted the NCF architecture~\cite{he2017neural} for our use case, in which the DL model receives a concatenation of three embeddings; the user embedding is analogous to alert embedding $a^{(m)}_{j,emb}$, and the item embedding is divided into two embeddings: \textit{(1)} a module embedding $m_t$, and \textit{(2)} a playbook graph embedding $pb^*$.
The model then returns a recommendation score indicating the suitability of the new playbook module $m_{t+1}$ that should be included in the playbook being created.
The range of the score is between zero and one, where the top recommendations receive higher scores.

\section{\label{sec:eval}Evaluation}

\subsection{Datasets}

\textbf{Evaluation platform:} In our experiments, for the SIEM platform, we used the Splunk Enterprise Security tool,\footnote{https://www.splunk.com/en\_us/products/enterprise\-security.html} and for the SOAR platform we used two different tools, Shuffle\footnote{https://shuffler.io/} and Splunk SOAR.\footnote{https://www.splunk.com/en\_us/products/splunk\-security\-orchestration\-and\-automation.html}

\noindent\textbf{Event logs datasets:} To evaluate the proposed method, we used two event logs datasets: Splunk Boss of the SOC (BOTS) v1/v2/v3 dataset\footnote{https://github.com/splunk/botsv1, https://github.com/splunk/botsv2, https://github.com/splunk/botsv3} and Splunk Attackdata (AD) dataset.\footnote{https://github.com/splunk/attack\_data}
These event logs were ingested into the Splunk SIEM.

\noindent\textbf{Alert rules:} For the BOTS dataset, we manually created analytical rules to generate alerts from the raw logs, and for the AD dataset, we exported the analytical rules from the Splunk Security Content (SSC) GitHub repository\footnote{https://github.com/splunk/security\_content} published by Splunk.
The Splunk CIM~\cite{cdm} schema was used to parse these alerts into a standardized format that can be used universally across any SIEM or SOAR platform. \\
The CIM schema includes 2,661 alert fields, i.e., keys, each representing a specific aspect of the alert.
To represent each alert in our recommender system, we converted the alert representation into a sparse vector format.
In this format, relevant keys in $K$ (i.e., keys with values that in not $\varnothing$) are labeled as one, while the remaining keys are labeled as zero, indicating their absence or irrelevance for a specific alert.
We extracted 55 types of alerts from the BOTS dataset using the manually created analytical rules.
From the SSC repository, we implemented 60 different endpoint-related analytical rules that trigger alerts in the AD dataset.

\noindent\textbf{Playbooks datasets:} We used a dataset consisting of playbooks crafted for the purpose of responding to alerts.
We used two different platforms to create playbooks, \textit{Shuffle SOAR} and \textit{Splunk SOAR}.
Using Shuffle SOAR, a total of 23 playbooks were created, each representing predefined sequences of modules to be executed in response to the alerts.
Similarly, using Splunk SOAR, we crafted 30 playbooks tailored for alert response. 
We used the following two generic playbook datasets published by Palo Alto Networks\footnote{https://github.com/demisto/content} and Splunk\footnote{https://research.splunk.com/} as our reference when creating the abovementioned playbooks.

\noindent\textbf{Alert rules -- Playbooks mapping datasets:} For our evaluation, we created three different mappings between the alert and playbook datasets.
We manually mapped the 55 alerts in the BOTS dataset to the 23 playbooks created using Shuffle SOAR; we refer to this mapping as $D_1$.
Similarly, we mapped 60 alerts from the AD dataset to 12 playbooks created using Splunk SOAR; we refer to this mapping as $D_2$.
For the third mapping $D_3$, we used all of the alerts in the BOTS dataset and mapped them to 18 playbooks created using Splunk SOAR.
A summary of these three datasets is presented in Table~\ref{tab:mapping_summary}.
Note that within each of the playbooks created, there is a designated start node, which serves as the entry point for the playbook upon receiving alert data and defines the starting point for the playbook's workflow.

\begin{center}
\begin{table*}[h]
\newcolumntype{C}[1]{>{\centering\arraybackslash}m{#1}}
\caption{Summary of the datasets used in our evaluation.}
\label{tab:mapping_summary}
% \resizebox{1.0\textwidth}{!}{%
\begin{adjustbox}{width=1.0\textwidth,center=\textwidth}
\begin{tabular}{@{}lllllllll@{}}
\toprule
\textbf{Mapping Name} & \textbf{Alerts Dataset} & \textbf{No. of Alert Rules} & \textbf{SOAR System} & \textbf{No. of Playbooks} & \textbf{No. of Modules Used} \\ \midrule
$D_1$ & BOTS & 55 & Shuffle  & 23 & 26 \\
$D_2$ & AD & 60 & Splunk & 12 & 21 \\
$D_3$ & BOTS & 55 & Splunk & 18 & 40 \\

\bottomrule
\bottomrule
\end{tabular}%
\end{adjustbox}
\end{table*}
\end{center}

\subsection{Experimental Setup}
To evaluate the performance of our recommender system, we used a 5-fold cross-validation approach based on the alert rules (hereafter will be referred to as alerts) in our datasets.
We split the alerts into five different subsets, each serving as a test set for the evaluation; in each case, the remaining subsets were used for training.

\begin{figure}[h]
    \centering
    \includegraphics[width=\linewidth,trim={0cm 0.2cm 0cm 0cm},clip]{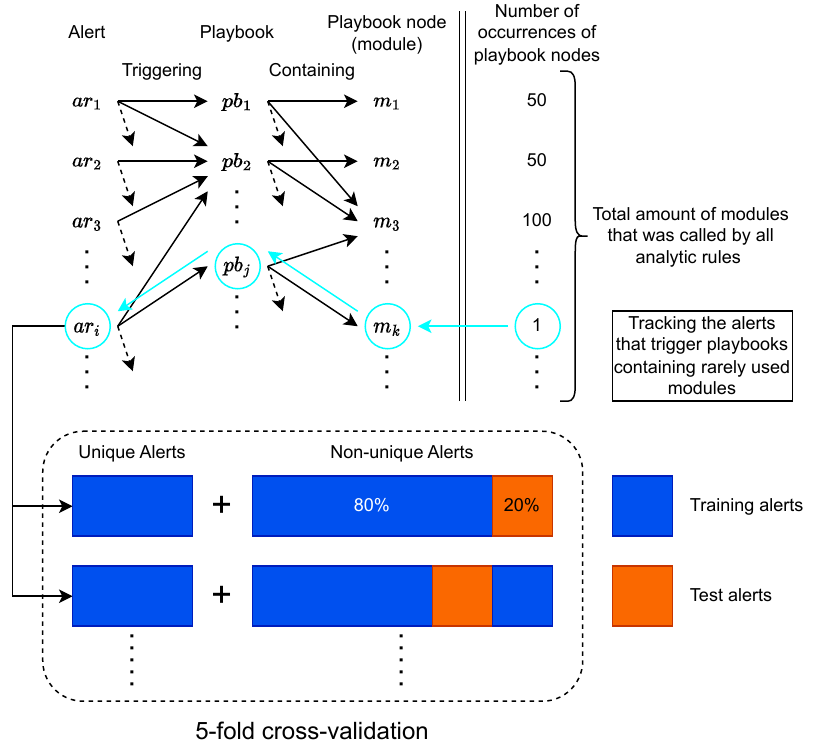}
    \vspace{-0.5cm}
    \caption{Alert data split based on the unique alerts and 5-fold cross-validation.}
    \label{fig:alert_split}
\end{figure}

To ensure that our training data includes all possible scenarios (specifically, all possible security modules), we identified a set of \emph{unique} alerts as follows (shown in Figure~\ref{fig:alert_split}).
First, we determine the total occurrence count of a specific module $m_i$ based on the mapping between alerts and playbooks. 
This count represents how many times module $m_i$ is invoked across all playbooks triggered by various alerts in the analytical rule repository.
Modules with low total counts (below four in our evaluation) are considered as unique.
Next, we identify the unique alert rules. 
Unique alerts are defined as those triggering playbooks that includes unique modules (i.e., performing rare actions).
The unique alerts were not included in the 5-fold cross validation split, and we added them manually to all training sets.
Note that the number of unique alerts in the datasets is relatively small, which is expected to have a limited impact on the conducted experiments.

\subsection{Recommendation Dataset Creation}
\label{Recommendation Dataset Creation}

\begin{figure*}[h!]
    \centering
    \includegraphics[width=0.99\textwidth]{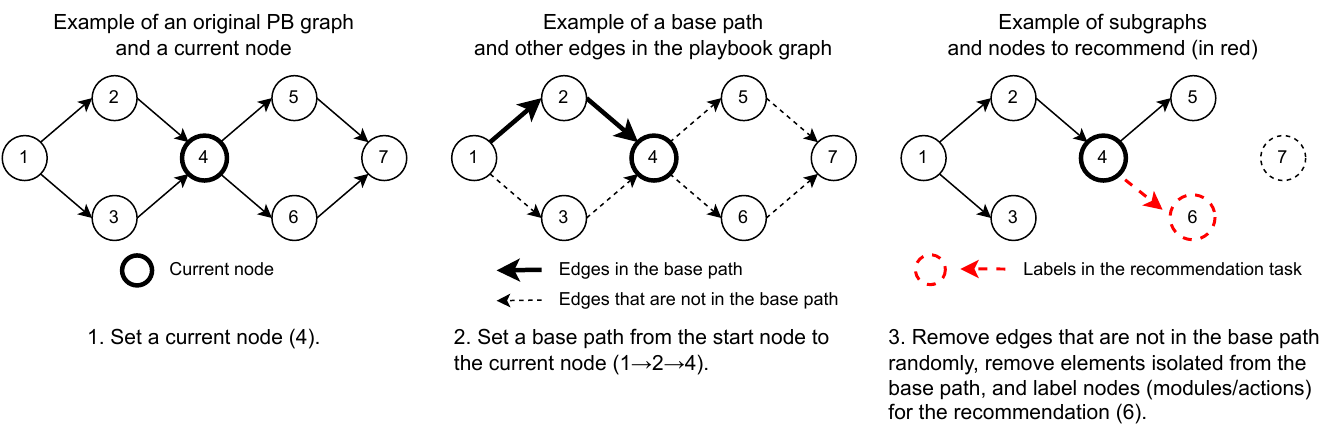}
    \vspace{-0.9cm}
    \caption{Playbook graph subtraction strategy used in recommendation dataset creation process.}
    \label{fig:subgraph_strategy}
\end{figure*}

To generate the module recommendation datasets based on realistic assumptions (for training the recommender system model and performing the evaluation), we repeated the following iterative procedure for each model inference and dataset:

\begin{enumerate}[noitemsep, nolistsep]
    \item Select an \textbf{alert rule} from the alert rule dataset.\label{step:alert}
    \item Select a \textbf{playbook} ($pb$) that can be triggered by the alert selected in step~\ref{step:alert}.
   
    \item Select a \textbf{current node} ($m_t$) in the playbook. 
    \item Create a \textbf{partial playbook} ($pb^*$) by maintaining a path from the beginning of the graph (start node) to the current node, $m_t$, and randomly removing edges from the other part of the playbook graph, as illustrated in Figure~\ref{fig:subgraph_strategy} (\textbf{context}). 
    The edges are pruned with 50\% probability, and isolated nodes are excluded.
    \item For each pair of partial playbook graph ($pb^*$) and current node ($m_t$), we create the \textbf{label vector}.
    The size of the label vector is set as the number of possible modules plus one (for the EOP artificial module).
    Each module in the label vector receives the value of one ('1') if it is connected from $m_t$ in the original playbook but not in the subgraph, and zero ('0') otherwise.
    The EOP module is only labeled as true when there are no modules to recommend from the current node in the subgraph.
    
\end{enumerate}

Consequently, our dataset consists of tuples of <alert rule, partial subgraph, current node, label vector>. 

\subsection{Model Configuration}

The training of \method starts with the training of the embedding modules, then the DL model follows, which receives the outputs of the embedding modules as the inputs.

\subsubsection{Embedding Models}
The architecture of the autoencoder used for the alert embedding is comprised of two layers with dimensions [2661, 256].
The ReLU activation function is applied to all layers except for the final layer, which uses a sigmoid function.
The resulting code size, which represents the embedding vector $a^{(m)}_{j,emb}$, is set at 16 dimensions.
The autoencoder is trained to minimize the binary cross-entropy loss with a learning rate of 0.1 for 2000 epochs.

For module embedding, we employ the node2vec\cite{grover2016node2vec} algorithm, using the Pytorch Geometric Project's~\cite{Fey/Lenssen/2019} implementation.
The node2vec model is trained by considering the probability of encountering other nodes in the neighborhood during random walks.
We carefully set the hyperparameters, such as walk length, context size, walks per node, P, Q, and the number of epochs, to control the exploration of neighbors and capture the relevant structural information of the unified playbook graph.
Specifically, we use the following hyperparameter values: embedding size - 16, walk length - four, context size - four, walks per node - three (the diameter of the graph minus one), P - five (to avoid repetition in random walks), and Q - 0.25 (to induce exploration outward in random walks), and we train the model for 10,000 epochs.

For playbook graph embedding, we employed the graph2vec~\cite{narayanan2017graph2vec} technique using the karateclub~\cite{karateclub} implementation for graph embedding.
The input for the graph2vec models consists of the original graphs of the playbooks in the training data.
We set the size of the graph embeddings at 16 to align with the alert and node embeddings.
We explored two approaches for playbook graph embedding representation -- without attributes and with attributes.
A graph representation without attributes only considers connections between a module and other modules.
In contrast, a graph representation with attributes includes the precise structure of a playbook in which nodes are associated with modules; therefore, multiple nodes representing the same module can be placed in the graph representation with attributes.

\subsubsection{DL Model Training}
In our experiments, we employed the DL model that consists of two hidden layers, each containing 64 units.
The DL model accepts an input vector of size 48, which is derived from the concatenation of three embeddings.
The training data is divided into random batches of 64 samples during the iterations.
The loss function used is binary cross-entropy, which is commonly employed in recommendation tasks, and the Adam optimizer was used for training with a learning rate of 0.001 for 1,000 epochs.

\subsection{Baselines}
We compared the proposed method's performance to that of two types of recommender systems methods: frequency-based recommendation and non-negative matrix factorization (NMF)~\cite{paatero1994positive,wang2012nonnegative}.

In frequency-based recommendation, i.e., popularity-based recommendation, frequently used items are suggested in the order of their frequency.
Although this is a lightweight and simple recommendation method, it represents a state-of-the-art method in the recommendation domain. 
NMF~\cite{paatero1994positive,wang2012nonnegative} is based on the idea of dimensionality reduction, and it can make use the context of other items in new item recommendations.

To create reasonable models for the recommendation in this realistic scenario, i.e., recommendation during the implementation, the baseline models were also trained using the same recommendation inference data (Section~\ref{Recommendation Dataset Creation}).
The frequencies for the frequency-based model were calculated from the true labels in the training data after 100 epochs.
This reflects the module appearance in the recommendation results, taking into account minor effects from randomization.
In the NMF model, both the existing modules from the current playbook graph and the true labels in the recommendation were treated as a single set of items, termed as latent factors. 
The state and recommendation results of one epoch in the training dataset were provided as the input for the NMF model, and the raw outputs were used as the recommendation scores in the evaluation.

\subsection{Results}

\begin{table*}[h]
\caption{Evaluation results.}
\label{tab:evaluation_results}
\begin{adjustbox}{width=1.0\textwidth,center=\textwidth}
\begin{tabular}{|l|l|cccc|cccc|cccc|}
\hline
\multicolumn{1}{|l}{} & \multicolumn{1}{|l}{} & \multicolumn{4}{|c}{Precision@k} & \multicolumn{4}{|c|}{Recall@k} & \multicolumn{4}{|c|}{Mean Average Precision@k} \\ \cline{3-14}
\multicolumn{1}{|l}{Dataset} & \multicolumn{1}{|l}{Model} & \multicolumn{1}{|c}{k=1} & \multicolumn{1}{c}{k=3} & \multicolumn{1}{c}{k=5} & \multicolumn{1}{c}{k=10} & \multicolumn{1}{|c}{k=1} & \multicolumn{1}{c}{k=3} & \multicolumn{1}{c}{k=5} & \multicolumn{1}{c}{k=10} & \multicolumn{1}{|c}{k=1} & \multicolumn{1}{c}{k=3} & \multicolumn{1}{c}{k=5} & \multicolumn{1}{c|}{k=10} \\ \hline \hline
D1 & Frequency-based & 0.5953 & 0.2746 & 0.1835 & \textbf{0.1022} & 0.5827 & 0.8063 & 0.8979 & \textbf{1.0000} & 0.5953 & 0.6936 & 0.7153 & 0.7304 \\
 & NMF-based & 0.0845 & 0.0492 & 0.0417 & 0.0342 & 0.0827 & 0.1444 & 0.2042 & 0.3345 & 0.0737 & 0.1025 & 0.1162 & 0.1345 \\
 & \method (without attributes) & 0.7950 & \textbf{0.3303} & \textbf{0.2040} & \textbf{0.1022} & 0.7782 & \textbf{0.9701} & \textbf{0.9982} & \textbf{1.0000} & 0.7887 & 0.8756 & 0.8826 & 0.8826 \\
 & \method (with attributes) & \textbf{0.8022} & 0.3291 & 0.2032 & \textbf{0.1022} & \textbf{0.7852} & 0.9665 & 0.9947 & \textbf{1.0000} & \textbf{0.7941} & \textbf{0.8768} & \textbf{0.8835} & \textbf{0.8835} \\
\hline
D2 & Frequency-based & 0.5171 & 0.2359 & 0.1625 & 0.0971 & 0.4895 & 0.6700 & 0.7690 & 0.9195 & 0.5171 & 0.5918 & 0.6154 & 0.6346 \\
 & NMF-based & 0.0096 & 0.0129 & 0.0197 & 0.0312 & 0.0090 & 0.0367 & 0.0933 & 0.2952 & 0.0078 & 0.0184 & 0.0305 & 0.0567 \\
 & \method (without attributes) & \textbf{0.7480} & \textbf{0.3372} & \textbf{0.2095} & \textbf{0.1056} & \textbf{0.7081} & \textbf{0.9576} & \textbf{0.9914} & \textbf{1.0000} & \textbf{0.7285} & \textbf{0.8479} & \textbf{0.8559} & \textbf{0.8573} \\
 & \method (with attributes) & 0.6891 & 0.3303 & 0.2080 & 0.1055 & 0.6524 & 0.9381 & 0.9848 & 0.9990 & 0.6711 & 0.8066 & 0.8180 & 0.8201 \\
\hline
D3 & Frequency-based & 0.5240 & 0.2447 & 0.1686 & 0.0995 & 0.5043 & 0.7064 & 0.8114 & 0.9577 & 0.5240 & 0.6094 & 0.6342 & 0.6526 \\
 & NMF-based & 0.0170 & 0.0167 & 0.0214 & 0.0229 & 0.0164 & 0.0481 & 0.1030 & 0.2204 & 0.0140 & 0.0281 & 0.0400 & 0.0549 \\
 & \method (without attributes) & \textbf{0.6290} & \textbf{0.3203} & \textbf{0.2030} & \textbf{0.1036} & \textbf{0.6054} & \textbf{0.9249} & \textbf{0.9769} & \textbf{0.9971} & \textbf{0.6183} & \textbf{0.7730} & \textbf{0.7848} & \textbf{0.7879} \\
 & \method (with attributes) & 0.6090 & 0.3200 & \textbf{0.2030} & \textbf{0.1036} & 0.5861 & 0.9240 & \textbf{0.9769} & \textbf{0.9971} & 0.5983 & 0.7624 & 0.7741 & 0.7773 \\
\hline
\end{tabular}
\end{adjustbox}
\end{table*}

%\begin{comment}
\begin{figure*}[htbp]
    \centering\includegraphics[width=0.80\linewidth]%,height=0.5\linewidth]
    {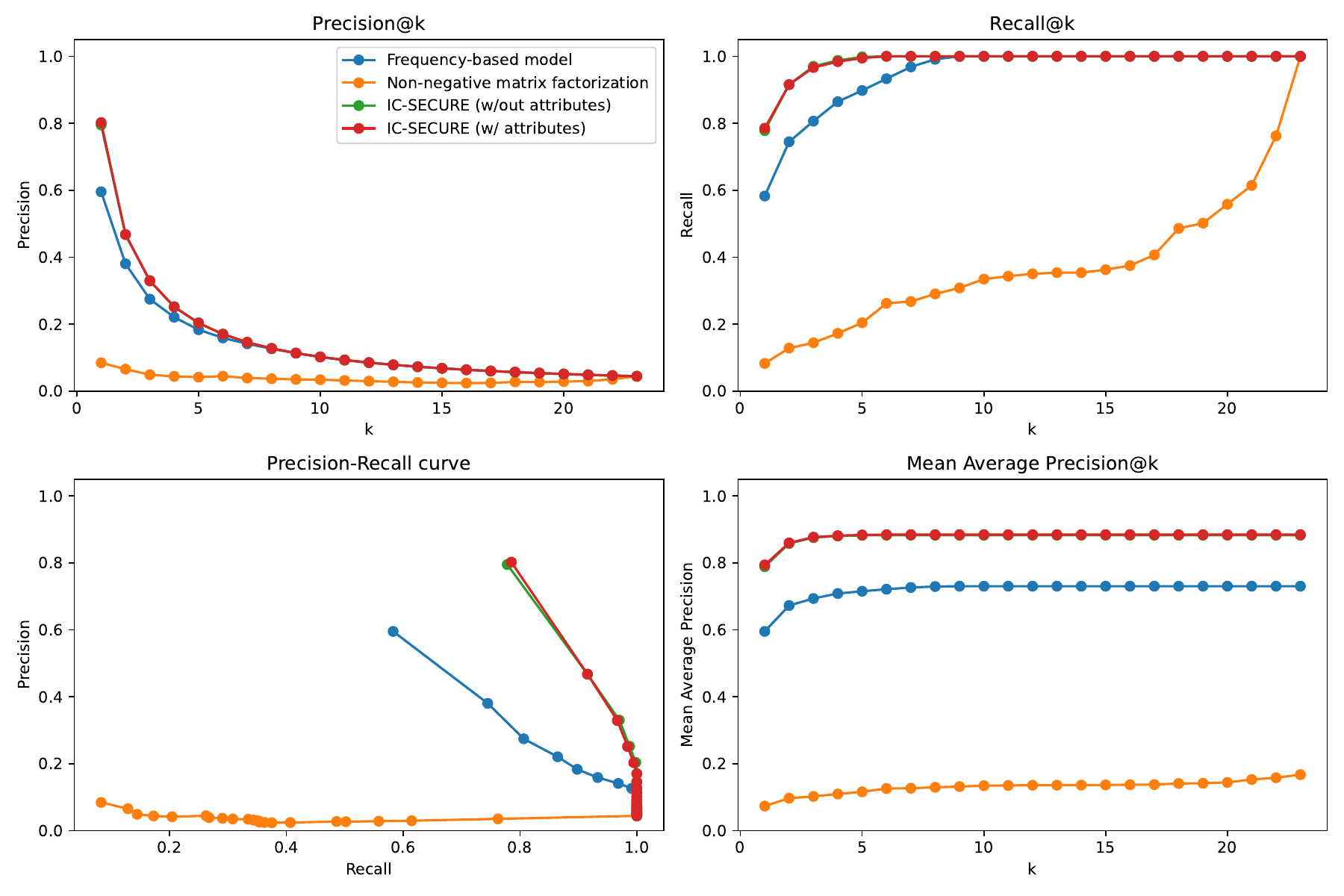}
    \vspace{-16pt}
    \caption{Evaluation results on dataset $D_1$.}
    \label{fig:d1_results_w_eop}
\end{figure*}

\begin{figure*}[htbp]
    \centering\includegraphics[width=0.8\linewidth]%,height=0.5\linewidth]
    {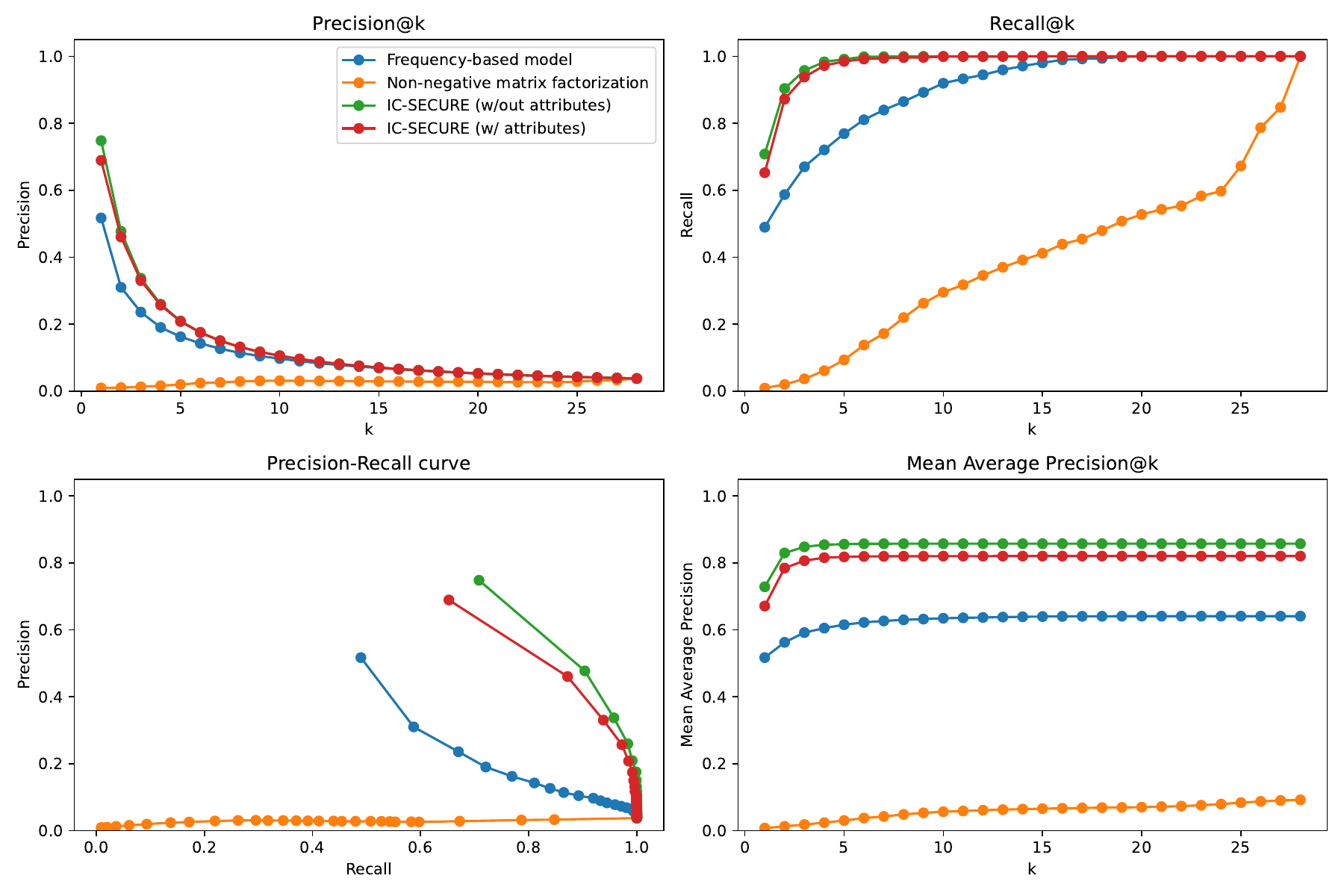}
    \vspace{-16pt}
    \caption{Evaluation results on dataset $D_2$.}
    \label{fig:d2_results_w_eop}
\end{figure*}

\begin{figure*}[htbp]
    \centering\includegraphics[width=0.8\linewidth]%,height=0.5\linewidth]
    {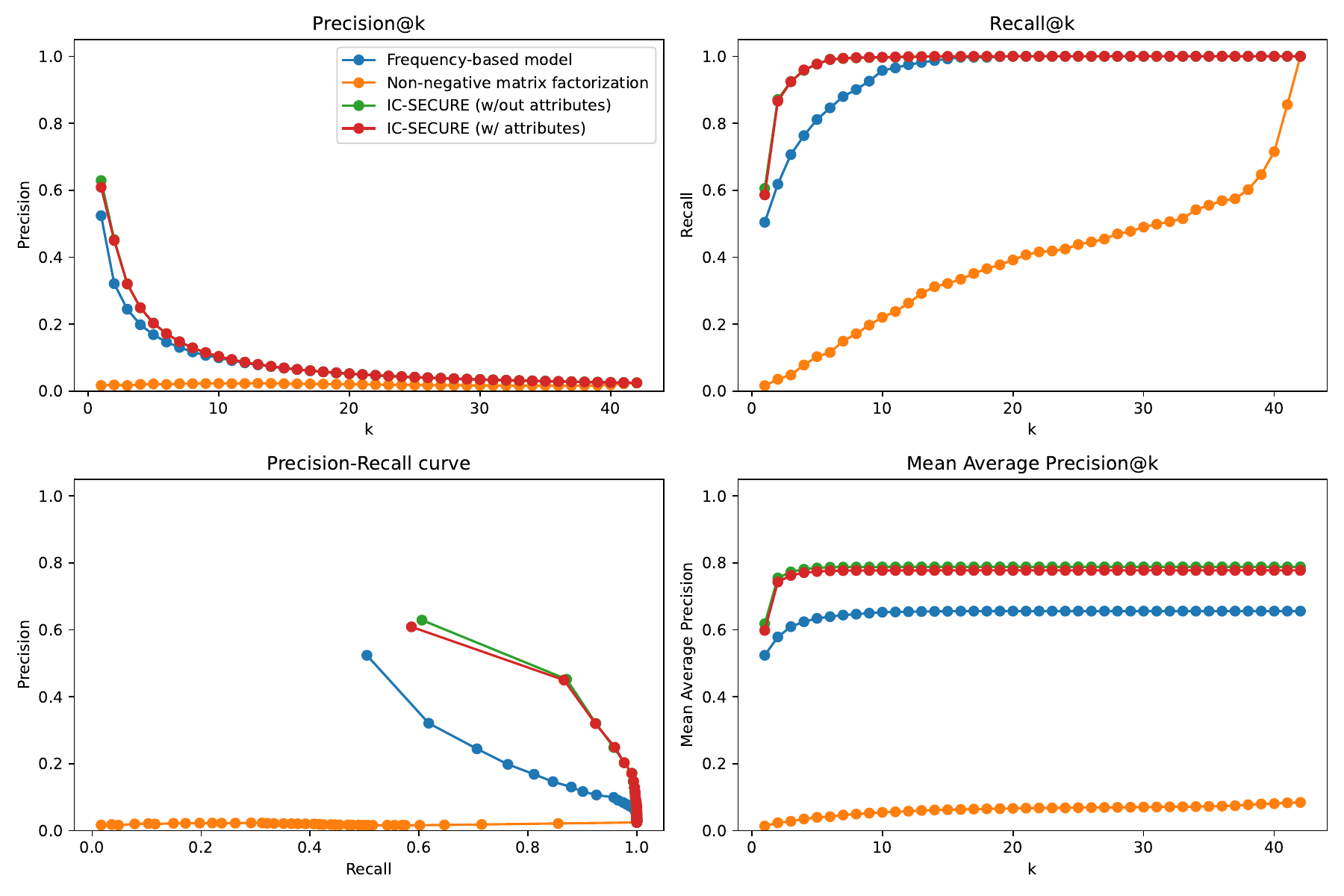}
    \vspace{-16pt}
    \caption{Evaluation results on dataset $D_3$.}
    \label{fig:d3_results_w_eop}
\end{figure*}
%\end{comment}
To assess the performance of the compared methods, we used the metrics of precision and recall at $k$, which denote the precision and recall values, when the recommendation model recommends the items with the top-$k$ scores. %, i.e., predicted as true values.
Here $k$ denotes the number of recommendations being provided by the model.
In the task of next module recommendation, $precision@k$ represents the probability that the first $k$ recommended modules are in the original playbook graph, and $recall@k$ represents the probability that the modules in the original playbook graph are in the first $k$ recommendations.

Figures~\ref{fig:d1_results_w_eop}-\ref{fig:d3_results_w_eop} and Table~\ref{tab:evaluation_results} present the precision, recall values and mean average precision (MAP) values for different $k$ on datasets $D_1$, $D_2$, and $D_3$, respectively.
Generally, the two variations of \method (with/without attributes in the graph embedding) outperform the baselines on all the metrics. 

%\textbf{Precision for small values of $k$.}
\textbf{Precise recommendations within smaller values of $k$}.
One important aspect in the multi-label recommendation is its precision in providing relevant recommendations for small values of $k$.
As can be seen from Table~\ref{tab:evaluation_results}, the precision values at $1$ of \method are higher than 0.6 (and in the range of 0.6-0.8) in all datasets, outperforming the baseline algorithms.
Additionally, the MAP values of \method are relatively high and stable within the range of 0.77-0.88 for $3 \leq k \leq 5$.
This indicates that \method is able to recommend the correct modules at the top of its recommendation list.

\textbf{Recommending all relevant items within smaller values of $k$}.
In addition, the recall values of \method for relatively small $k$ $(3 \leq k \leq 5)$ are above 0.95 in all datasets, again, outperforming the baseline algorithms.
This means that in more than 95\% of the cases, all the correct modules were covered in the top five recommendations.
We believe that $k=5$ is a reasonable and practical setting to be used in real SOAR GUI applications for playbook implementation.

\textbf{Effect of items corpus size on recommendations}.
Regardless of the model being evaluated, we have seen that the $precision@k$ values decline as the corpus of items (i.e., modules) increases.
When \method provides less number of recommendations i.e., $(3 \leq k \leq 5)$, the precision and MAP values remain high across all experimental datasets.
This indicates that \method is capable of providing relevant recommendations for small values of $k$ and strengthens our argument that $k=5$ is a practical setting for model's application in module recommendation for playbook creation task.

\textbf{Graph embedding with/without attributes}.
Finally, we compare the results of \method when employing graph embedding with attributes and without.
Intuitively, graph embedding with attributes should capture a more comprehensive representation of the playbook graphs, leading us to anticipate improved recommendations. 
However, as depicted in Figures~\ref{fig:d1_results_w_eop}-\ref{fig:d3_results_w_eop} and Table~\ref{tab:evaluation_results}, both configurations exhibit comparable performance.

\subsection{Demonstration}

\begin{figure*}[h]
    \centering
    \includegraphics[width=1.0\linewidth]{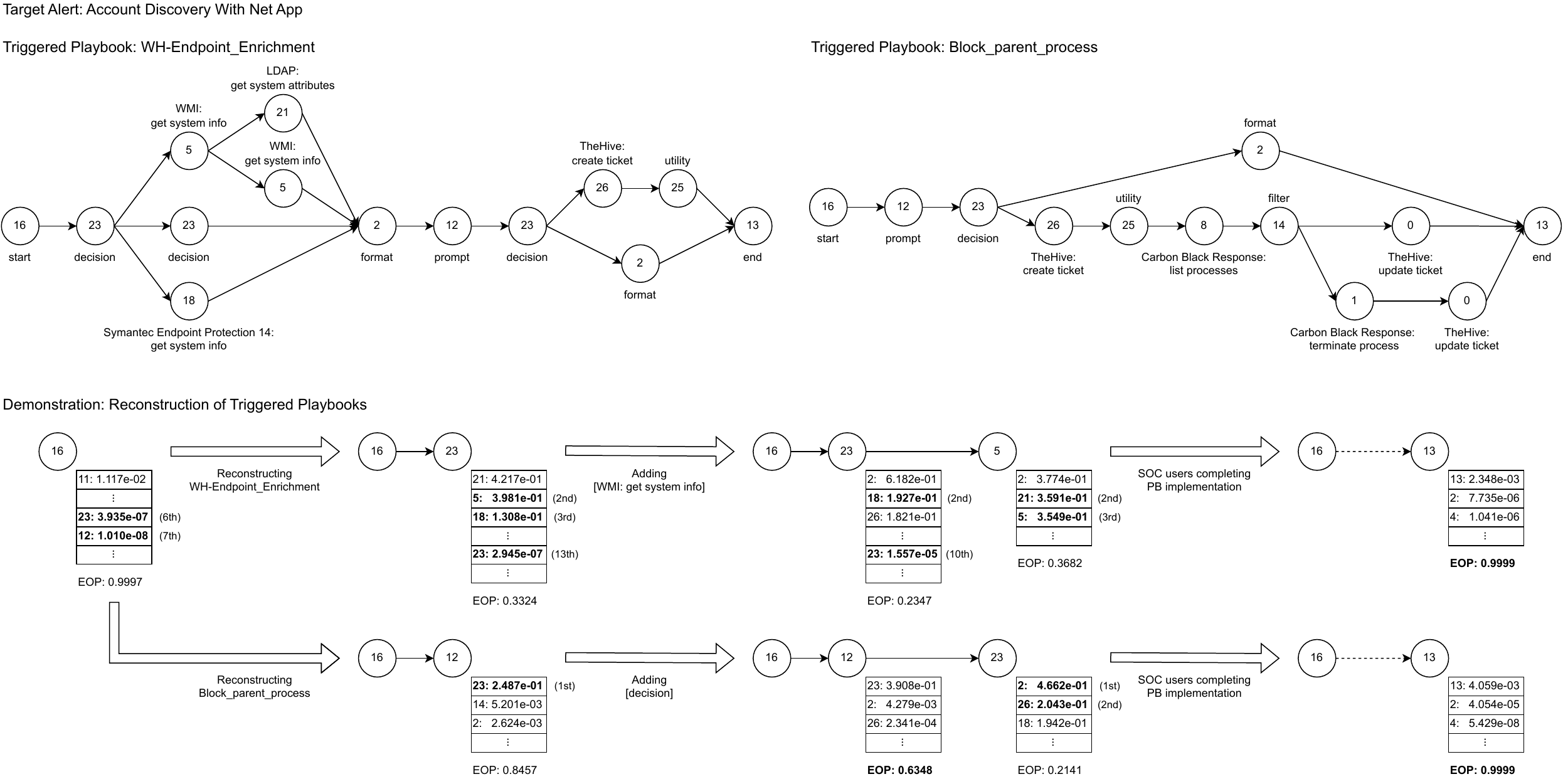}
    \caption{\method demonstration of recommendation tasks with the dataset $D_2$. 
    Two playbook graphs on the top are the graphical images of the playbooks triggered by an alert (Account Discovery With Net App). 
    The flow below shows the results of the \method recommendation with different contexts of graphs and current nodes.
    Each list of values contains recommendation scores, along with the module IDs, i.e., "\textbf{Module ID: Recommendation score}", with the situation of the graph of the current playbook implementation and the current node in the graph, and the scores of EOP below. 
    The modules that should be recommended appear in bold, along with the rank of the recommendation.}
    \label{fig:demonstration_d2}
\end{figure*}

In Figure~\ref{fig:demonstration_d2} we demonstrate \method's ability to make recommendations in a realistic scenario.  
Dataset $D_2$ is used in this simulated scenario, along with an alert called "Account Discovery With Net App," which \emph{was not} part of the data used to train the recommender model.
In this scenario, the model should be able to recommend appropriate modules depending on the context, which leads to reconstructing the following playbooks: "WH-Endpoint\_Enrichment" and "Block\_parent\_process."
Regarding the recommendation results with the context near the initial state of a playbook graph, where the graph contains a small number of nodes, the ranks for the modules that should be recommended are not high.
On the other hand, in most cases with rich context, the model succeeds in recommending the appropriate modules to reconstruct those playbooks with higher ranks.
This is more significant in the attempt to reconstruct a response playbook (Block\_parent\_process).
This demonstration also illustrates the \method's ability to stop making recommendations in cases in which the EOP value is relatively high and recommendation values for other modules are relatively lower.

\section{\label{sec:conclusion}Conclusion}
In this paper, we proposed a novel recommender system, \method, aimed at providing security module recommendations for playbooks in SOAR systems.
The module recommendations provided by \method assist security analysts in their efforts to create new playbooks to respond to novel unknown alerts. 
In \method, a deep neural network model based on neural collaborative filtering technique serves as the recommender system that takes as input the context in the form of alert features and the current state of the incomplete playbook, along with the information about propagating node in the playbook, and provides relevant module recommendations for the subsequent nodes in the playbook being created.
The evaluation results demonstrate \method's ability to understand the context and make use of it to successfully recommend the appropriate security modules, outperforming the frequency-based and NMF baseline methods.

In future work, we plan to extract additional features from the input and output parameters of each module to be included in the playbook and also from the conditions on the links connecting these modules that provide additional context to the model in providing more relevant recommendations during the playbook creation task.
Furthermore, we aim at extending our method to support complete end-to-end pipeline where, once an alert is received by the SOAR system, a DL-based model handles the alert and deploys appropriate responses automatically (i.e., dynamically and autonomously creating on-the-fly playbooks) and thus reducing the burden on security analysts.

%-------------------------------------------------------------------------------
\newpage
\clearpage
\bibliographystyle{IEEEtran}
\bibliography{main}

%%%%%%%%%%%%%%%%%%%%%%%%%%%%%%%%%%%%%%%%%%%%%%%%%%%%%%%%%%%%%%%%%%%%%%%%%%%%%%%%
\end{document}